\documentstyle[preprint,tighten,aps]{revtex}
\begin{document}
\begin{center} 
{\bf \large The fermionic contribution to the spectrum
of the area operator in nonperturbative quantum gravity}\\
\vspace*{10pt}

{\bf Merced Montesinos$^{a,b}$\footnote{E--mail: 
merced@fis.cinvestav.mx} and Carlo Rovelli$^a$\footnote{E--mail: 
rovelli@pitt.edu}}\\
$^a$Department of Physics and Astronomy, University of 
Pittsburgh,\\
Pittsburgh, PA 15260, USA.\\

$^b$Departamento de F\'{\i}sica, Centro de Investigaci\'on y de 
Estudios Avanzados del I.P.N.,\\ 
Av. I.P.N. No. 2508, 07000 Ciudad de M\'exico, M\'exico.
\end{center}

\begin{center}
{\bf \small Abstract}
\end{center}
The role of fermionic matter in the spectrum of the area operator 
is analysed using the Baez--Krasnov framework for quantum 
fermions and gravity.  The result is that the fermionic 
contribution to the area of a surface $S$ is equivalent to the 
contribution of purely gravitational spin network's edges tangent 
to $S$.  Therefore, the spectrum of the area operator is the same 
as in the pure gravity case.\\

PACS: 04.60.Ds, 04.20.Cv.
\baselineskip 20pt

Loop quantum gravity \cite{Loop}, the nonperturbative approach to 
quantum gravity, is nowadays a mathematically well-defined theory 
with a powerful {\it predictive} character (see \cite{Review} for 
a recent review). The theory is based on the Hamiltonian 
formulation of general relativity due to Ashtekar 
\cite{VariablesA} which, as was shown in \cite{VariablesB}, is 
the ADM formulation of the (self--dual sector of the) Plebanski 
action \cite{VariablesC}.  At present, the theory is usually 
formulated in terms of the real $SU(2)$ Ashtekar connection, 
whose use has been advocated by Barbero \cite{VariablesD}, and 
which can be obtained through a canonical transformation from the 
original complex Ashtekar variables.  Amongst the most striking 
results of loop quantum gravity are the spectra of the area and 
volume operators \cite{Area,al,Volume}, and the computation of the 
entropy of black holes \cite{Black}.  These results point to the 
existence of discrete aspects of spacetime at the Planck length 
$l_P =\sqrt{ G\hbar/c^3}$.
 
The research in loop quantum gravity is presently developing 
along three main directions.  The first of these focuses on the 
physics of black holes \cite{Black,Black2}.  The second deals 
with the the Hamiltonian constraint \cite{Hamiltonian,Pullin} and 
with Feynman--type formulations \cite{Feynmann} of the quantum 
dynamics.  The third studies the coupling of matter fields to 
quantum gravity.  For instance, in \cite{Masa} the contribution 
of the quantum states to the fermionic mass has been studied, 
while the possibility of a quantum-gravity induced vanishing of 
the cosmological constant has been explored in 
\cite{Cosmological}.  Although these results are preliminary, 
they indicate that unexpected phenomena may result from the 
coupling between quantum matter fields and the non-perturbative 
quantum gravitational field.  This paper deals with this third 
direction.  In particular, our aim is to study the modifications 
of the spectrum of the area operator due to the presence of a 
fermion field.

Our main result is that the spectrum of the area is not 
altered by the presence of fermions.  This result was already 
anticipated in \cite{al}, on the basis of general considerations 
on the form of gravity-matter theories.  Here, we verify it 
explicitly within a specific fermionic model.
 
The model we consider is the Einstein--Weyl theory.  We follow 
the approach developed in \cite{Krasnov}, which has its roots in 
the previous works of \cite{Hugo}, and we take (as in 
\cite{Krasnov}) SU(2) as the relevant internal group.  For our 
purposes, the key difference from other matter couplings is that 
in the case of the spin-1/2 field there is a matter contribution 
to the gravitational Gauss constraint.  The other matter fields 
do not have this property.  This difference is the reason for the 
nontrivial interplay between the area and the fermions.
 
The area operator ${\widehat A}_{S}$ associated to a 
two--dimensional surface $S$ is well 
understood \cite{Area,al}. Let us
recall its definition. We take $S$ to be an open surface 
with boundaries (we will 
comment on closed surfaces later on). The operator can be written 
as
\begin{equation} 
{\widehat A}_{S}  =  8 \pi \beta l^2_P \ 
\sum_{v\in S} \sqrt{{\widehat A}^2_{S,v}}, 
\end{equation}
where $\beta$ is the Immirzi parameter\cite{Parameter} and 
${\widehat A}^2_{S,v}$ is the vertex area operator.  This 
operator acts on the vertices $v$ of the spin network states 
\cite{Networks}, lying on the surface $S$.  It can be written as
\begin{equation}
\label{Aver}
{\widehat A}^2_{S,v}  =  \frac{1}{2} \epsilon^{AC} \epsilon^{BD} 
\left( 
{\widehat J}^{(d)}_{AB} - {\widehat J}^{(u)}_{AB} 
\right) 
\left(
{\widehat J}^{(d)}_{CD} - {\widehat J}^{(u)}_{CD}
\right).    
\end{equation}
Here ${\widehat J}^{(u)}= \sum_i {\widehat J}^{(u)}_i 
\,^{outgoing} + \sum_j{\widehat J}^{(u)}_j \,^{incoming}$ and 
${\widehat J}^{(d)}=\sum_i {\widehat J}^{(d)}_i \,^{outgoing} + 
\sum_j{\widehat J}^{(d)}_j \,^{incoming}$ are the (symmetric) 
`link operators'\footnote{These operators are the `spinorial 
version' of the Ashtekar--Lewandowski (see \cite{al}) angular 
momentum operators.} associated with the edges of the spin 
network on the two sides (`up' and `down') of the surface 
$S$; the sums run over the edges $\gamma_i$, `outgoing' and 
`incoming', at the vertex.  For a particular edge $\gamma_j$, 
the action of the operator ${\widehat J}_{j\,\,AB}$ on a a 
cylindrical function $\Psi_{\Gamma,f}(A)$ (see \cite{Networks}) 
is given by 
\begin{eqnarray}
{\widehat J}_{j\,\,AB} \Psi_{\Gamma,f}(A) :=  
\left\{
\begin{array}{ll}
& +\frac{i}{2} \epsilon_{C(A} U_{B)}\,^D [\gamma_j, A] 
\frac{\partial f}{\partial U_C\,^D [\gamma_j,A]},\quad 
\mbox{if $\gamma_j$ is `outgoing',}\\
& -\frac{i}{2} U_{C(A} [\gamma_j,A] \delta_{B)}\,^D
\frac{\partial f}{\partial U_C\,^D[\gamma_j,A]},\quad
\mbox{if $\gamma_j$ is `incoming',}\\
\end{array}
\right.
\end{eqnarray}
The above formulae follow directly from the definition of the 
area operator, and are valid with as well as without fermions.

In the pure gravity case, the gravitational Gauss constraint 
${\widehat G}[N]^{Einstein} \Psi_{\Gamma,f} (A)=0$, implies that 
at every vertex of the spin networks lying on the surface $S$ the 
following condition holds
\begin{eqnarray}
{\widehat J}^{(u)}_{AB} +{\widehat J}^{(d)}_{AB} +
{\widehat J}^{(t)}_{AB} = 0,
\end{eqnarray}  
where ${\widehat J}^{(t)}_{AB}$ is the sum of the link operators 
associated to the edges that exit $v$ tangentially with respect to 
$S$ (neither `up' nor `down').  This fact allows us to 
rewrite (\ref{Aver}) as
\begin{eqnarray}
{\widehat A}^2_{S,v} &  = &  \frac{1}{2} \epsilon^{AC} \epsilon^{BD} 
\left ( 
2 {\widehat J}^{(u)}_{AB} {\widehat J}^{(u)}_{CD} + 
2 {\widehat J}^{(d)}_{AB} {\widehat J}^{(d)}_{CD}
- {\widehat J}^{(t)}_{AB} {\widehat J}^{(t)}_{CD}
\right ). 
\end{eqnarray}
Since the three terms of the sum can be diagonalized 
simultaneously, the purely gravitational spectrum of the area
\begin{equation}
	A_S =  8 \pi \beta l^2_P \ 
\sum_{v} \sqrt{\frac{1}{2} j^{u}_{v}(j^{u}_{v}+1)+
\frac{1}{2} j^{d}_{v}(j^{d}_{v}+1) - 
\frac{1}{4} j^{t}_{v}(j^{t}_{v}+1)}, 
\label{spectrum} 
\end{equation}
follows easily.  Here $j^{u}_{v}$, $j^{d}_{v}$ and $j^{t}_{v}$ 
are the total spins of the upgoing, downgoing and tangential 
edges of the vertex $v$.  

Let us come now to the Einstein--Weyl theory.  Let $\eta^A$ and 
${\widetilde \pi}_B$ be the fermionic field and its conjugate 
momentum, respectively. The purely gravitational Gauss constraint 
$G_{AB}^{Einstein}$ becomes \cite{Gauss}
\begin{eqnarray}
{\cal G}_{AB} (x) := G_{AB}^{Einstein}(x) +
\eta_{(A}(x) {\widetilde\pi}_{B)}(x). 
\label{Gaussfer}
\end{eqnarray}  
A straightforward calculation shows that the quantum version of 
this constraint on a fermionic cylindrical function 
$\Psi_{\Gamma,f}(A,\eta)$ \cite{Krasnov} 
implies that the following condition must hold at every vertex of the 
spin networks which lies on the surface $S$
\begin{eqnarray}
{\widehat J}^{(u)}_{AB} + {\widehat J}^{(d)}_{AB} + 
{\widehat J}^{(t)}_{AB} + {\widehat S}_{AB} & = & 0\, ,
\label{Condition1}
\end{eqnarray}
where
\begin{eqnarray}
{\widehat S}_{AB}(v) \cdot \Psi_{\Gamma,f}(A,\eta) = \left\{ 
\begin{array}{ll}
0 & \mbox{, if no fermions sit in the vertex $v$,}\\ 
-\frac{i}{2} \eta_{(A} \frac{\partial^L}{\partial \eta^{B)} (v)} 
\Psi_{\Gamma,f}(A,\eta) & \mbox{, if fermions sit in the vertex 
$v$.}
\end{array} 
\right.\label{Condition2}
\end{eqnarray}
(The superscript `L' stands for left derivatives 
\cite{Henneaux}).  Taking into account equation 
(\ref{Condition1}), the formula for the square of the vertex area 
operator (\ref{Aver}) can be written as
\begin{eqnarray}
{\widehat A}^2_{S,v} &  = &  \frac{1}{2} \epsilon^{AC} \epsilon^{BD} 
\left [ 
2 {\widehat J}^{(u)}_{AB} {\widehat J}^{(u)}_{CD} + 
2 {\widehat J}^{(d)}_{AB} {\widehat J}^{(d)}_{CD}
-\left ( {\widehat J}^{(t)}_{AB} + {\widehat S}_{AB}
\right )\cdot \left ( {\widehat J}^{(t)}_{CD} + 
{\widehat S}_{CD} \right ) \right ]\, .
\label{areafer}
\end{eqnarray}
We analyze here this operator, i.e. its action on the fermionic 
spin network states\cite{Krasnov}.\\

First of all, notice that owing to the Grassman property of the 
fermionic field, there are restrictions on the number of 
fermionic fields which can sit at a given vertex of the fermionic 
spin networks.  In general, for gauge--invariant fermionic spin 
networks, there can be only zero, one or two fermions at each 
vertex of the fermionic spin networks.  That is, the quantum 
state can only be independent, linearly dependent, or 
quadratically dependent, from the value of the fermion field in a 
given vertex.  Therefore, there are three cases in the analysis 
for the action of the ${\widehat A}^2_{S,v}$ operator.\\

I. No fermions at the vertex.  From equations (\ref{Condition2}), 
the meaning of (\ref{areafer}) is clear: if there are no 
fermionic excitations at a given vertex of the spin network, 
${\widehat S}_{AB}$ vanishes, and the action of ${\widehat 
A}^2_{S,v}$ operator reduces to the pure gravitational one.  This 
result comes from the fact that the action of the vertex area 
operator is `local', i.e.  the contribution of fermionic spin 
networks and pure gravitational spin networks is exactly the same 
for every vertex which has no fermionic excitations.\\

II. Two fermions at the vertex.  When there are two fermions in a 
given vertex of the fermionic spin network, a straightforward 
calculation shows that the action of the operator $ -\frac{1}{2} 
\epsilon^{AC} \epsilon^{BD} \left ( {\widehat J}^{(t)}_{AB}+ 
{\widehat S}_{AB} \right )\cdot \left ( {\widehat J}^{(t)}_{CD}+  
{\widehat S}_{CD} \right )$ in (\ref{areafer}) on the two 
fermions $\eta^E \eta^F$ which sit at the vertex vanishes.  This 
result follows from the anticommuting properties of the spinor 
field, together with the symmetry of the ${\widehat J}^{(t)}_{AB}$ 
operator.  Therefore the contribution to ${\widehat A}^2_{S,v}$, 
in this case, comes only from the edges of the spin network, i.e.  
even though there are two fermions in the vertex, the action of the 
fermionic operator ${\widehat S}_{AB}$ is missing.  Consequently, 
the contribution in this situation is as in (\ref{areafer}) with 
${\widehat S}_{AB}=0$.\\

III. One fermion at the vertex.  It is instructive to consider 
first a simplified case (a `one-side' vertex).  Assume that a 
vertex has no `down' nor tangential links, but only `up' 
links.  Thus ${\widehat J}^{(u)}\neq 0$, ${\widehat J}^{(d)}=0$, 
${\widehat J}^{(t)}=0$.  In the pure gravitational case, such a 
vertex has vanishing contribution.  Assume a fermion sits in the 
vertex.  We have   
\begin{equation}
2 \epsilon^{AC}\epsilon^{BD} {\widehat S}_{AB} {\widehat 
S}_{CD}\,\,\eta^E = \left ( \frac12\right ) \left 
(\frac12+ 1 \right )\,\,\eta^E,
	\label{spin12}
\end{equation}
(that is, the fermion has spin $\frac12$).  Using this and 
(\ref{Condition1}) we have that (\ref{areafer}), reduces to
\begin{eqnarray}
{\widehat A}^2_{S,v} &  = & \frac14 \left (\frac12 \right ) 
\left ( \frac12 + 1 \right )\, .
\end{eqnarray}
Therefore the fermion contributes to the area formula 
(\ref{spectrum}).  The contribution is the same as the one from 
a tangential link with spin $j=\frac12$.  As a second instructive 
example, consider the case of a vertex with no tangential edges 
and a single fermion.  Using (\ref{spin12}), we obtain the contribution to 
the square of the vertex area operator
\begin{eqnarray}
{\widehat A}^2_{S,v} &  = &  \frac{1}{2} \epsilon^{AC} \epsilon^{BD} 
\left [ 
2 {\widehat J}^{(u)}_{AB} {\widehat J}^{(u)}_{CD} + 
2 {\widehat J}^{(d)}_{AB} {\widehat J}^{(d)}_{CD}
\right ] 
- \frac14 \left ( \frac12\right ) \left ( \frac12 + 1 \right ). 
\label{averferB}
\end{eqnarray}   
Again, the fermionic contribution is the same as the contribution 
of a tangential link with spin $\frac12$. In general, in fact, it 
is easy to see from the fact that the `tangential' angular 
momentum ${\widehat J}^{(t)}_{AB}$ and the fermion's `spin' 
${\widehat S}_{AB}$ in (\ref{areafer}), appear only in the 
combination ${\widehat J}^{(t)}_{AB} + {\widehat S}_{AB}$, that 
the two cases above illustrate the general situation: the effect 
on the area operator of the presence of a fermion is the same as 
the effect of the presence of an additional tangential edge with 
spin $j=\frac12$.\\

In conclusion, we have studied the contribution of the spin-1/2 
matter field (within the framework of the Einstein--Weyl theory) to 
the spectrum of the area operator.  Our result is that the effect 
of the fermions at a given vertex of the fermionic spin 
network--lying on the surface $S$-- is equivalent to the effect 
of a tangential spin network edge with spin $j=\frac12$ at such 
vertex.  Therefore, the spectrum of the area operator does not 
change if fermionic matter is present, in spite of the fact that 
the quantum states do.

Our result can be understood intuitively as follows.  The area 
operator `sees' only the edges of the spin network that are not 
tangential to the surface (see Eq.(\ref{Aver})).  This is due to 
the fact that the loop states (of which the edges of the spin 
network are composed) carry {\it transversal\/} quanta of area, 
contributing only to the area of surfaces that are not tangent to 
the loop itself.  See \cite{Ro93} for a discussion of the 
relevant geometry.  Thus, from the point of view of the area 
operator, a tangential spin network edge is like 
an edge `moving out of the 
manifold' carrying angular momentum with itself.  Now, precisely 
the same is true for fermions: in fact, it has been repeatedly 
noticed that in loop quantum gravity a fermion behaves precisely 
as a gravitational loop `continuing out of the manifold'.  In 
\cite{Hugo}, it was noticed that the dynamics of a fermion is the 
same as the dynamics generated by the {\it purely 
gravitational\/} quantum hamiltonian constraint on the end point 
of loop states `cut open', more precisely, generated by the 
shift operator \cite{Loop}. In \cite{Smolin}, it was even 
suggested to interpret this very surprising fact in terms of John 
Wheeler's ideas of particles as wormholes: a fermion is the 
point in which a gravitational line of flux plunges into a small 
wormhole.  In \cite{BaezPiriz} the similarity between fermions 
and end points of loop states was extended to the loop's end 
points on boundaries of the manifold.  Here, we have shown that a 
fermion behaves as `a gravitational line of flux continuing out 
of the manifold' also as far as the area is concerned.

The case of a {\it closed} surface is slightly different, 
due to the `fermion number conservation' discussed in detail in 
\cite{al}. This is a restriction of the spectrum which can be 
easily understood using the `old' overcomplete loop basis 
instead of the spin network basis (see \cite{Networks}): every 
loop that enters the closed surface must exit it, so that a closed 
surface is always crossed by an {\it even} number of loops.  
Since a fermion plays the role of end-point of a loop, this 
restriction disappears in the presence of fermions.  Therefore, 
the theory with fermions differs from the pure gravity theory in 
the fact that, in the presence of fermions, the spectrum of the 
area of a closed surface is the same as the spectrum of the area 
of an open surface \footnote{We thank Jerzy Lewandowski for this 
observation.}.

We notice that since the classical limit of the area observable 
is presumably recovered in the $j\rightarrow \infty$ limit, the 
contribution of the fermionic spin network states, which is 
given by spin $j=1/2$ terms only, can be seen as a {\it pure 
quantum} effect.  On the other hand, we recall that the 
contribution of the spin $j=1/2$ terms plays a dominant role for 
the value of the entropy of black holes \cite{Black}; therefore 
the results presented here might have implications for black hole 
entropy.

The case of the Einstein--Dirac \cite{Gauss} system follows 
the same lines as the present case if one follows (as here) the 
formalism of \cite{Krasnov}; the only difference with respect to 
the present case is the contribution of two types of fermionic 
operators, associated with the two two--component fermionic fields. 
From the general results in \cite{al}, the same result also holds
for a general class of matter fields including Thiemann's 
fermions \cite{Thiemann}.  On the other hand, the role of the 
fermionic matter in the spectrum of the volume operator 
\cite{Volume} is not yet known, and deserves to be studied.

\section*{Acknowledgments} MM thanks Carlo Rovelli for many 
hours of joint computations and for teaching him deep issues of 
quantum gravity.  E-mail correspondence from Kirill K Krasnov and 
Alejandro Corichi about fermionic spin networks and area 
operators respectively is especially acknowledged. We are also 
very grateful to Jerzy Lewandowski for important 
observations and for clarifying the relation between our work and 
reference \cite{al}. MM thanks all the 
members of the relativity group of the 
Department of Physics and Astronomy of the University of 
Pittsburgh for their warm hospitality.  MM's postdoctoral 
fellowship is funded through the CONACyT of M\'exico, fellow number 
91825.  Also MM thanks support from the {\it Sistema Nacional de 
Investigadores} of CONACyT. This work has been partially 
supported also by NSF Grant PHY-95-15506.

\end{document}